\documentclass[journal]{vgtc}         

\ifpdf
  \pdfoutput=1\relax                   
  \usepackage{graphicx}                
  \DeclareGraphicsExtensions{.pdf,.png,.jpg,.jpeg} 
\else
  \usepackage{graphicx}                
  \DeclareGraphicsExtensions{.eps}     
\fi%

\graphicspath{{figures/}{pictures/}{images/}{./}} 

\usepackage{microtype}                 
\PassOptionsToPackage{warn}{textcomp}  
\usepackage{textcomp}                  
\usepackage{mathptmx}                  
\usepackage{times}                     
\usepackage{cite}                      
\usepackage{tabu}                      
\usepackage{booktabs}                  
\usepackage{amsmath}
\usepackage{subfig}
\usepackage{diagbox}
\usepackage{boldline}
\usepackage{url}
\usepackage[bookmarks=false]{hyperref}
\usepackage{xcolor} 
\usepackage{tikz}

\usepackage{float}

\onlineid{1049}

\vgtccategory{Research}
\vgtcpapertype{Technology}

\title{The Security-Utility Trade-off for Iris Authentication and Eye Animation for Social Virtual Avatars}

\author{Brendan John, \textit{Student Member, IEEE}, Sophie J{\"o}rg, Sanjeev Koppal, \textit{Senior Member, IEEE}, and Eakta Jain}
\authorfooter{
\item
 Brendan John is a PhD student at the University of Florida. \\ E-mail: brendanjohn@ufl.edu.
\item
 Dr. Sophie J{\"o}rg is an Associate Professor at Clemson University. \\ E-mail: sjoerg@clemson.edu.
\item
 Dr. Sanjeev Koppal is an Assistant Professor at the University of Florida.  E-mail: sjkoppal@ece.ufl.edu.
 \item
 Dr. Eakta Jain is an Assistant Professor at the University of Florida. \\ E-mail: ejain@cise.ufl.edu
}

\shortauthortitle{John \MakeLowercase{\textit{et al.}}: The Security-Utility Trade-off for Iris Authentication and Eye Animation for Conversational Virtual Agents}

\abstract{The gaze behavior of virtual avatars is critical to social presence and perceived eye contact during social interactions in Virtual Reality. Virtual Reality headsets are being designed with integrated eye tracking to enable compelling virtual social interactions. This paper shows that the near infra-red cameras used in eye tracking capture eye images that contain iris patterns of the user. Because iris patterns are a gold standard biometric, the current technology places the user's biometric identity at risk. Our first contribution is an optical defocus based hardware solution to remove the iris biometric from the stream of eye tracking images. We characterize the performance of this solution with different internal parameters. Our second contribution is a psychophysical experiment with a same-different task that investigates the sensitivity of users to a virtual avatar's eye movements when this solution is applied. By deriving detection threshold values, our findings provide a range of defocus parameters where the change in eye movements would go unnoticed in a conversational setting. Our third contribution is a perceptual study to determine the impact of defocus parameters on the perceived eye contact, attentiveness, naturalness, and truthfulness of the avatar. Thus, if a user wishes to protect their iris biometric, our approach provides a solution that balances biometric protection while preventing their conversation partner from perceiving a difference in the user's virtual avatar. This work is the first to develop secure eye tracking configurations for VR/AR/XR applications and motivates future work in the area.
} 


\CCScatlist{ 
 \CCScat{K.6.1}{Management of Computing and Information Systems}%
{Project and People Management}{Life Cycle};
 \CCScat{K.7.m}{The Computing Profession}{Miscellaneous}{Ethics}
}

\renewcommand{\manuscriptnotetxt}{\textcopyright2020 IEEE. Personal use of this material is permitted.  Permission from IEEE must be obtained for all other uses, in any current or future media, including reprinting/republishing this material for advertising or promotional purposes, creating new collective works, for resale or redistribution to servers or lists, or reuse of any copyrighted component of this work in other works. \\
   DOI: \href{https://ieeexplore.ieee.org/abstract/document/8998133}{10.1109/TVCG.2020.2973052}}

\vgtcinsertpkg

\newcommand{\tvar}[1]{\mathrm{#1}} 

\newcommand{\mytilde}{\raise.17ex\hbox{$\scriptstyle\mathtt{\sim}$}}

\begin{document}

\maketitle

\section{Introduction}
Eye tracking will transform virtual and mixed reality. Major hardware companies are integrating eye trackers into head-mounted displays\,(HMDs) to enable applications ranging from intuitive gaze-based interfaces\,\cite{pai2017gazesphere,rajanna2018gaze,zhang2019accessible}, foveated rendering\,\cite{patney2016towards,bastani2017foveated}, and streaming optimization\,\cite{lungaro2018gaze,feng2019viewport}. Foveated rendering is driving eye tracking within VR headsets due to the potential to both optimize resources and reduce simulator sickness\,\cite{patney2016towards,sun2017perceptually}. For social virtual reality with hyper-realistic virtual avatars\,\cite{lombardi2018deep,macquarrie2019perception}, eye tracking is required to transfer non-verbal social cues from the user to his or her conversational virtual avatar.



Because of the networked nature of social platforms and the use of cloud-based rendering techniques for VR\cite{mueller2018shading}, it is expected that XR devices will follow an `always on and connected' model. Streaming eye tracking data makes it susceptible to attacks. Most critically, the iris image of the user is vulnerable. The iris image is a gold standard biometric that is used in high security applications, such as border customs\cite{iris_netherlands_customs}, and is recognized as such by headset manufacturers\cite{hololens2}.
John et al.\cite{John2019} showed that typical eye tracker eye images, if stolen, could be used to biometrically identify as a user. They presented a proof of concept solution that blurred the eye image to remove the high frequency patterns that form each person's unique iris signature. They evaluated this solution for a target viewing task. However, for such a solution to be impactful, it is also necessary to determine the consequences of a security mechanism for specific applications. We focus on the application of eye tracking to animate the eyes of virtual avatars, as eyes are critical to realism and naturalness of avatars, gaze is a crucial social cue in conversations, and inadvertently altering a user's gaze may result in unintended changes in how he or she is perceived.
 
Our first contribution is to discuss the theoretical basis of this problem and a proposed solution, provide a novel hardware mechanism to achieve the solution, and evaluate its ability to reduce accuracy of iris authentication. Our second contribution is to determine detection thresholds for the amount of image defocus that can be applied before a difference in eye animations is perceived. Our third contribution is a study to determine how image defocus impacts perceived eye contact, attentiveness, naturalness, comfort, and truthfulness of the conversational avatar. Based on this work, it is possible to recommend to a user how to create their preferred level of security for eye tracking, and how much impact this setting will have on the perceived characteristics of their virtual avatar. More broadly, this work motivates the need to investigate the security-utility tradeoff for a wide range of XR applications and develop eye tracking configurations that prioritize security.

\section{Background}

\textbf{Eye Tracking in Virtual Reality} \,\, Current applications of eye movements in VR are driving investments in the next generation of eye tracking hardware. Applications include foveated rendering\cite{patney2016towards,bastani2017foveated}, which optimizes computational resources in rendering by reducing resolution in the periphery, streaming algorithms that reduce the bandwidth of streamed 360$^\circ$ content\cite{lungaro2018gaze,hu2019sgaze,feng2019viewport}, intuitive interfaces for navigation and predicting intent\cite{pai2017gazesphere,bednarik2012you}, subtle gaze direction using luminance cues in the periphery to guide attention\cite{8446215}, redirected walking methods that take advantage of saccadic masking and blinks to orient the user within a limited physical space\cite{sun2018towards,langbehn2018blink}, classifying neurodegenerative disease through eye movements\cite{orlosky2017emulation}, virtual experiences designed to improve joint attention of children with ASD\cite{mei2018towards}, and modeling how users explore 360$^\circ$ content\cite{sitzmann2018saliency}. Eye tracking hardware in VR ranges from video-based oculography\cite{kassner2014pupil}, electro-oculography\,(EOG)\cite{bolte2015subliminal}, photo-sensor oculography\,(PS-OG)\cite{zemblys2018developing,li2017ultra}, and magnetic sclera coils\cite{whitmire2016eyecontact}. EOG, PS-OG, and sclera coil eye trackers provide gaze estimation without imaging the eye itself, however video-based eye trackers are the most readily available solutions today. EOG and sclera coil approaches are invasive, as they require electrodes to be attached to the user's head or a magnetic contact lens to be worn by the user. PS-OG trackers are still being evaluated in terms of power usage and the ability to deploy within consumer devices, as the current implementation occludes the user's field of view\cite{zemblys2018developing}. Companies like Facebook, HTC, and Magic Leap have opted for a non-invasive video-based eye tracker that captures images of the eye, including the iris and other identifiable features like eyebrows\cite{dong2011eyebrow}. Thus, there is a need to investigate techniques that secure the iris during gaze estimation.

\textbf{Eye Movements for Conversational Virtual
Avatars} \,\, Eye movements play an important role in non-verbal communication, and thus are critical in creating compelling social interactions with virtual avatars. For example, Steptoe et al.\cite{steptoe2010lie} showed that the presence of eye movements caused participants to more accurately determine if an avatar was being truthful or not when compared to an avatar without eye movements. This is important for conversational avatars that discuss sensitive information, such as medical diagnoses\cite{volonte2018empirical}. The animation of virtual eyes can be data-driven or generated by procedural algorithms that model the dynamics of the eye. Realistic eye animations may include characteristics such as micro-saccadic jitter, blinks, eye lid displacement, and pupil diameter\cite{ruhland2014look}. Results from Duchowski et al.\cite{duchowski2015eye} suggest that data-driven eye animations are perceived as more natural than procedural animations. J{\"o}rg et al.\cite{jorg2018perceptual} found that subtle variations in the amplitude of noise within data-driven eye animations influenced how natural the animations were perceived. This suggests that a small amount of spatial noise in the signal may be detected, and have a negative impact on the naturalness of eye animations. Results from Garau et al.\cite{garau2001impact} suggest that a virtual avatar rendered with naturalistic eye and head movements did not improve communication over an audio-only conversation, when the eye and head movements do not match the context of the conversation. The authors also showed that an avatar with eye movements based on the current conversation produced similar responses in attentiveness and involvement to that of a video call with a real person. This implies that while models can be used to generate natural eye movements for an avatar, they may not contain the non-verbal cues and subtleties needed to simulate a real conversation. In these cases real eye tracking data is critical. In this paper we focus on data-driven eye movements in the absence of cues like blinks, eyelid movement, or pupil dilation to isolate the influence of perceived gaze direction and dynamics of the eyeball. 

\textbf{Privacy \& Security in Eye Tracking} \,\, There is a growing concern in keeping eye movement data private and secure in both real-time applications\cite{liebling2014privacy}, and published datasets\cite{liu2019differential}. Publicly available datasets release de-identified gaze data from individuals viewing VR videos\cite{david2018dataset}, the social interactions of children with ASD\cite{duan2019dataset}, and individual responses to emotional content such as nude imagery and faces\cite{ramanathan2010eye}. Sensitive information, such as personality traits\cite{hoppe2018eye} and neurological diagnoses\cite{leigh2015neurology}, could be linked to individuals that contributed to the aggregate data. To protect against this type of attack, differential privacy techniques have been proposed for securing heatmaps and other gaze-based features\cite{liu2019differential,steil2019privacy}. However, they are constrained to dealing with already recorded gaze data and not real-time streams. 

Mobile eye trackers rely on videos from an eye camera that captures the user's eye, and a front facing scene camera that records what they see. The scene camera is akin to wearable devices that are always on and recording video data. Public perception of these devices is overwhelmingly negative, as seen with the initial release of the Google Glass, as they infringe on the privacy of both the user and bystanders\cite{denning2014situ,naeini2017privacy,rashidi2018you}.  Daily users of eye tracking technology trade-off the privacy of their everyday actions for the benefit of activity logging, gaze-based interfaces, and assistive applications\cite{hoyle2014privacy,ahmed2016addressing,venuprasad2019characterizing,mei2018towards}. Steil et al. have developed a privacy approach specifically for the scene camera, using a controlled shutter to disable the video feed in private situations\cite{steil2019privaceye}. The eye camera is unique in that it captures raw eye movements and personally identifying information without any layer of security. Previous approaches for wearable-based privacy and security do not apply to this context. This paper focuses on a solution to protect against unauthorized iris-based identification from eye images.

\textbf{Iris Authentication} \,\, Infrared images of the eye with sufficient resolution capture iris patterns unique to the individual. Iris recognition places in the top tier of biometrics as it is universal, distinct, permanent, and robust against spoofing attacks\cite{jain2004biometricrecognition}. It is important to keep the iris pattern secure, as recognition methods are robust to poor lighting\cite{kahlil2010generation}, off-axis imaging\cite{daugman2007new}, occlusion\cite{daugman2009iris}, and distance\cite{algawhari2018iris}, making the biometric accessible at times when the user may not consent. Iris authentication has been long established through the work of John Daugman\cite{daugman1993high} and many others\footnote{Please see\cite{gale2014review} for a review.}, as a statistically valid method for recognition of an individual. As a result, iris patterns have been trusted for identification at voting booths\cite{iris_somali_election}, border customs\cite{iris_netherlands_customs}, schools\cite{iris_kenya_schools}, and in hospitals\cite{iris_hospitals}. These applications highlight the sensitivity of information that could be accessed if a hacker is able to steal identity through a biometric.
Thus, the presence of a user's iris within a dataset or application places the user's identity at risk.


\textbf{Defocus-based Identity Preservation} \,\, Rana and colleagues presented a systems argument for why applications that process images and videos do not necessarily need access to the raw image feed\,\cite{jana2013scanner,jana2013enabling}. Neustaedter et al.\cite{neustaedter2006blur} explored adding blur to increase privacy of a tele-conference video feed. They found that there is no general purpose blur level that preserves utility across all scenarios in this context. For example, the participants specified a much higher amount of blur in video that captured embarrassing activities such as picking their nose or changing clothes, compared to daily computer work. Participants were asked to identify the activities being performed in each video, with the level of blur being decreased until they could confidently classify the activity. The computed blur thresholds and classification rate determine that blur is effective at increasing privacy while retaining utility, but that the trade-off must be evaluated across applications and sensors. Hasan et al.\,\cite{hasan2018viewer} investigated various image filters such as masking, blurring, and pixelation with respect to their effectiveness in obscuring specific features of the content as well as retaining the utility and aesthetics of the photograph. They reported that blur was effective at obscuring the gender of the photographed person, though not so much the ethnicity or expression. Ultimately they determined that there is no `one size fits all' solution for every scenario, and object size or security context can influence the optimal method. Pittaluga and Koppal\cite{pittaluga2015privacy} have implemented a similar blur-based privacy approach within the context of micro-scale image sensors. A hardware-based approach is used to add blur, as opposed to a software-based Gaussian blur. The use of optics to scatter light before the image is captured creates blur on the camera sensor. Applications like head tracking, person tracking, and facial recognition are explored with several types of camera sensors\,(thermal, IR , RGB) imaging the user. Each camera configuration and application must be optimized and designed to balance the trade-off between security and utility. Our work investigates adding blur to eye images pre-capture, however the goal is to do so without modifying the stock hardware or optics. This allows consumers to control their own privacy, as current consumer technology would lack any specialized privacy hardware. 

John et al.\cite{John2019} have proposed the only existing method to protect the iris biometric within eye tracking images. Gaussian blur is applied to the eye images to remove high frequency details from iris patterns. A monocular glasses-based eye tracker was used to collect data from five participants, in which eye images were captured and matched to each other. The authors found that an eye camera at 320x240 resolution is able to capture iris patterns that successful identify each individual without false positives. Their results suggest that a Gaussian blur with $\sigma=5$ pixels is needed to reach the highest level of privacy, where no frames from any individual could be correctly recognized. Utility for the collected data was determined by an on-screen target viewing task, where blur at $\sigma=5$ produced gaze error of less than 1.5$^\circ$ visual angle. At higher levels of blur pupil detection rates drastically decrease to 60\% on average, resulting in a gaze data stream with gaps and low confidence values. Applying blur in software creates a risk that image data could be compromised before security is enforced. Our contribution to the state of the art is to explore the theoretical basis of a defocus-based solution, and propose and evaluate a hardware instantiation that is compatible with a popular eye tracker design. Importantly, we investigate the security-utility tradeoff of defocus parameters when utility is defined as how a virtual avatar is perceived, rather than data-level numerical error as in John et al.\cite{John2019}.

\section{Security Vulnerability and Solution}
\label{sec:iris_security}
The newest wave of VR and AR devices include integrated eye tracking devices, and are susceptible to identity theft and spoofing attacks. In this section we describe the threat model that puts the user at risk, and propose defocus as a solution to enable secure eye tracking configurations. We provide a theoretical basis of a solution and evaluate it with respect to degrading the iris biometric and errors in gaze estimation while viewing on-screen targets.   

\subsection{Threat Model}
Iris patterns present in eye tracking data streams can serve as a password, and are continuously streamed when an eye tracker is in use. This data stream is subject to a man-in-the-middle attack if images are sent over a network. In configurations where images are not streamed over a network, they are still subject to attacks when data is transferred at the hardware level.

An approach to protecting eye images is to only stream gaze data that is relevant to the application\cite{tobii_privacy_policy}. Image data is encapsulated within a processing unit, reducing the chance that a malicious user can gain access. However, this also restricts applications that may utilize the iris for improved gaze estimation\cite{Chaudhary_Pelz_2019}, realistic rendering of the user's eye\cite{francois2009image}, and iris authentication in cases where it is desired, such as logging into the Microsoft Hololens 2. 

Beyond risks in how manufacturers handle this sensitive data, the user must also control the permissions for third party applications that may access gaze and eye image data. There are growing concerns over how companies, large and small, handle sensitive data, with Facebook having their largest security breach most recently in 2018. A simple approach that alleviates all of these risks is to remove iris features prior to the image being recorded by the camera sensor. Then, even if a hacker gains access to the images they will not be able to use the iris patterns for authentication. Our work takes this approach.


\begin{figure}[!hb]
    \centering
    \includegraphics[width=\linewidth]{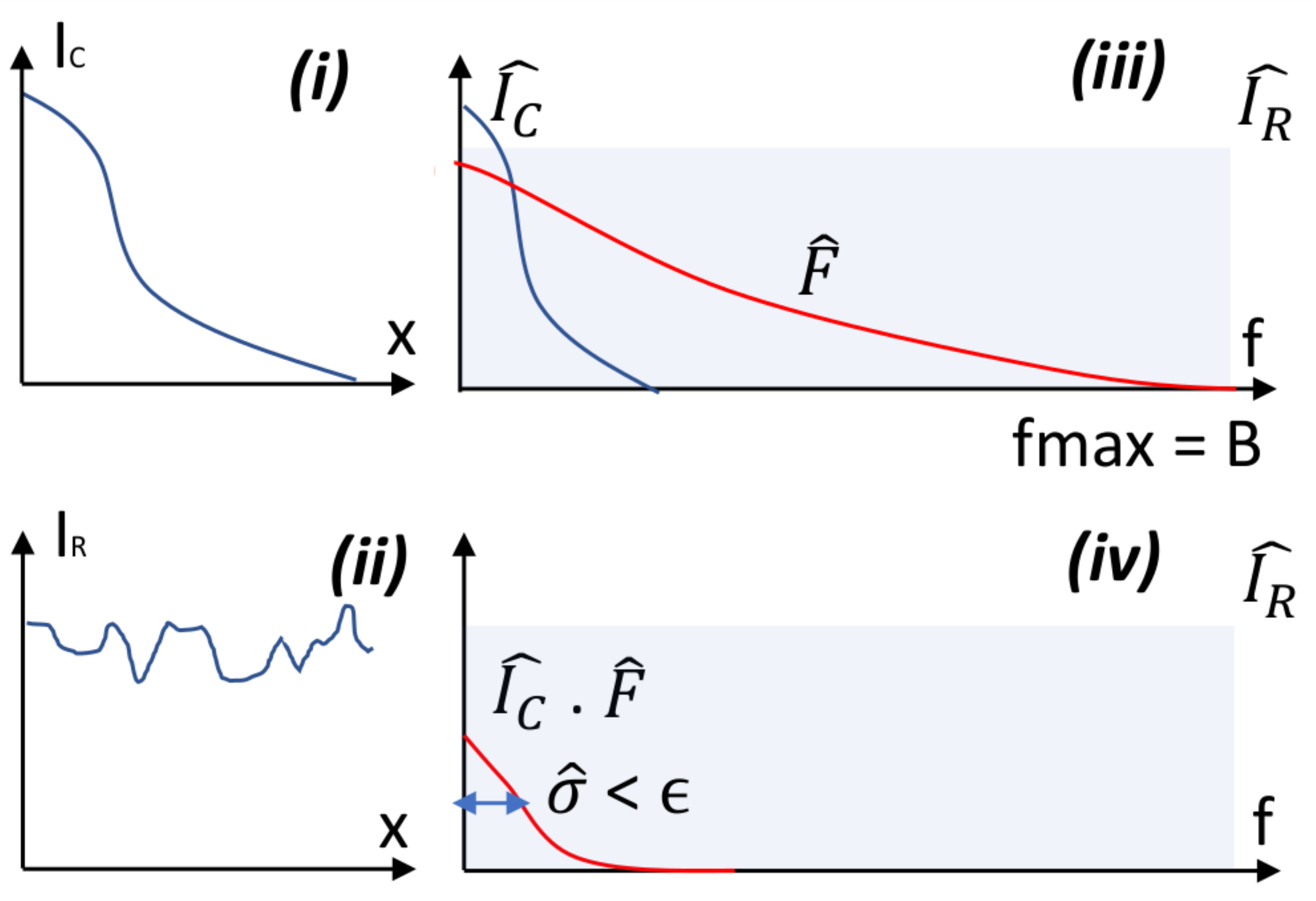}
    \caption{In (i-ii) we depict, in 1D, the eye tracking image component $I_C$ and the iris texture component $I_R$. In (iii) we show these again in the frequency domain, where the filter $F$ is depicted as Gaussian blur and $\hat{I}_R$ is some distribution spanning the shaded region. In (iv) we see $I_C$ remains usable for eye tracking when the standard deviation $\hat{\sigma}$ from $\hat{I}_C\cdot\hat{F}$ is not less than a minimum detectable bound $\epsilon$.}
    \label{fig:spatial_and_frequency_domain}
\end{figure}

\subsection{Frequency-based argument for proposed solution}
\label{sec:proposed_blur}
Our main assumption is that the eye tracking signal component $I_C$ and the iris texture component $I_R$ are \emph{separable}, and the image can be expressed as $I = I_C +I_R$. Without loss of generality, we assume these are 1D functions, but all our arguments below hold for 2D signals. $I_C$ is the component of the image that contains the eye tracking signal, and is modeled with a Gaussian distribution $I_C\approx N(\mu=0,\sigma_C)$. This Gaussian disk captures eye tracking features, such as corneal reflection highlights or pupil extent\cite{hansen2009eye}. The other component, $I_R$, corresponds to iris texture. Although iris textures have broad variation, we can assume that $I_R$ is band-limited, since the highest frequency in the signal is limited by image resolution. Let us denote the largest possible frequency as $B$.

While $I_C$ contains primarily low frequency content\,(defined by the standard deviation $\sigma_C$), $I_R$ contains both low and high frequency content, with the higher frequencies being the identifying features\,(up to the maximum frequency $B$). Figure\,\ref{fig:spatial_and_frequency_domain} illustrates the general functional form in spatial and frequency domain for these two signals. 

Consider a low-pass filter $F$ that is convolved with the image. We consider optical defocus, where the filter form is $F(x)$=$N(\mu$=$0,\sigma)$, i.e., a Gaussian blur. When $I$ is convolved with $F(x)$, the result is
\begin{equation}
    I_D(x) =  I(x) \ast F(x) \\
           = I_C(x) \ast F(x) + I_R(x) \ast F(x)\\
           = I'_C(x) + I'_R(x).
           \label{eq:applyblurconv}
\end{equation}

\noindent \emph{Our claim:} It is possible to select $F(x)$ such that the eye tracking features are still detectable in $I'_C$, while $I'_R$ no longer contains the higher frequencies that enable iris-based authentication. Let $\hat{I}_C$ and $\hat{I}_R$ denote the Fourier transform of $I_C$ and $I_R$ respectively, and $\hat{F}$ denote the Fourier transform of $F$. Note that the Fourier transform of a Gaussian distribution is also a Gaussian with standard deviation $\hat{\sigma}=\frac{1}{\sigma}$, i.e., $\hat{F} \approx N(\mu=0,\hat{\sigma}_F)$ and $ \hat{I}_C\approx N(\mu=0,\hat{\sigma}_C)$. In Fourier domain,  Eq.\ref{eq:applyblurconv} is
\begin{equation}
   \hat{I}_D(x) = \hat{I}_C(x)\cdot\hat{F}(x) + \hat{I}_R \cdot\hat{F}(x).
   \label{eq:applyblurconv2}
\end{equation}
\noindent \emph{Upper bound (i.e. how much defocus is too much):} If $\hat{\sigma}_F < \hat{\sigma}_C$, then the first term of Eq.\,\ref{eq:applyblurconv2} comprises the multiplication of two zero mean Gaussian functions, which yields a Gaussian function with zero mean and $\hat{\sigma}<<\hat{\sigma}_C$.  In the spatial domain, this corresponds to a Gaussian with $\sigma>>\sigma_C$. Intuitively, the corneal highlight or pupil extent has been heavily blurred, leading to difficulty in gaze location estimation. In frequency terms this means that the disk in the image is blurred enough that its transformed Gaussian dual has a indiscernible standard deviation, i.e. $\hat{\sigma} < \epsilon$, where $\epsilon$ is vanishingly small. Thus, this imposes an upper limit on $\sigma_F$.  

\noindent \emph{Lower bound (i.e. how much defocus is too little):} Within the second term of Eq.\,\ref{eq:applyblurconv2}, which contains the texture information necessary for iris-based authentication, the higher frequencies have been attenuated as a result of the point-wise multiplication with a Gaussian that has fallen off. Since $\hat{I}_R$ is band-limited by maximum frequency $B$, it's extent is within the range covered by $\hat{F}$. If $\sigma_F$ is large enough the values of $\hat{F}$ are extremely small and ``zero out'' the values of $\hat{I}_R$ during point-wise multiplication. This imposes a lower limit on $\sigma_F$, such that the identifying features of the iris are removed from $I(x)$.

%
\textbf{Our proposed approach:} Optical defocus is produced by increasing the distance between the camera and the user's eye, forcing the iris region out of focus. While some configurations allow the camera to be adjusted or even feature a lens with adjustable focus, eye trackers with limited access to the camera may require additional optics or hardware to be installed. In our configuration the amount of defocus is controlled by varying the distance between the eye and the camera.


\begin{figure*}[!ht]
    \centering
    \includegraphics[width=\linewidth]{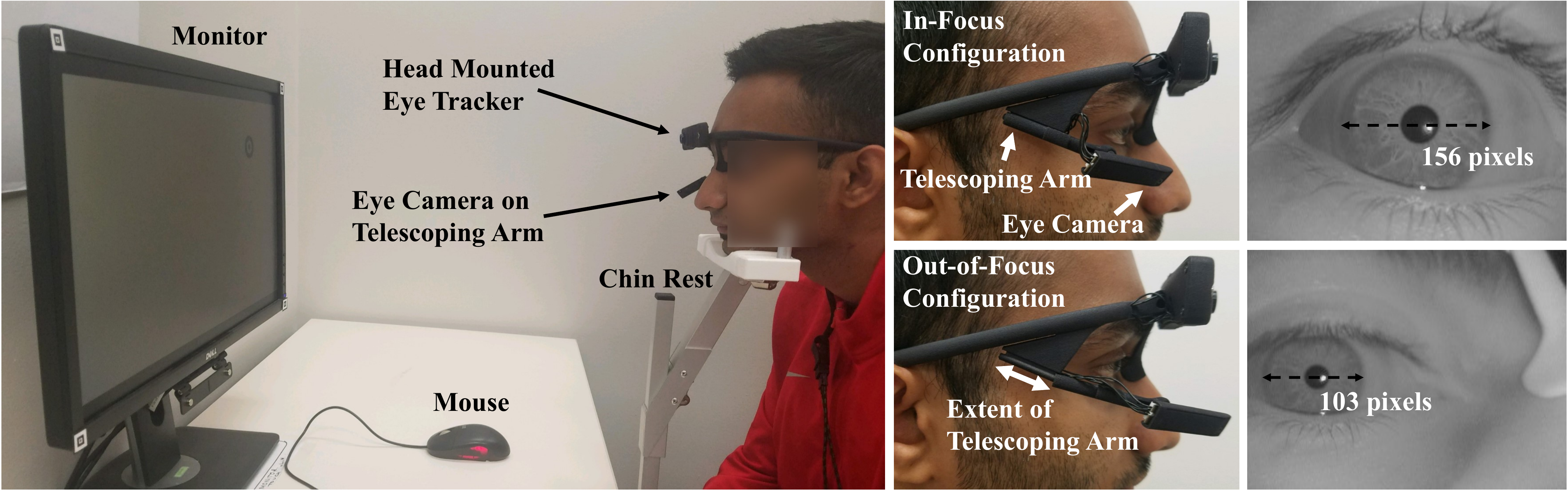}
    \caption{Experimental setup for evaluation. The adjustable telescoping arm of the eye tracker is used to create an out-of-focus configuration. Eye images from an in-focus configuration\,($23.3$mm) and out-of-focus configuration\,($35.4$mm) are shown.}
    \label{fig:exp_setup}
\end{figure*}

\subsection{Implementation}
\label{sec:auth_implement}
We implement optical defocus to create a secure eye tracking configuration. We use a Pupil Labs Pro glasses-based eye tracker with an adjustable telescoping arm to increase camera distance. Example eye images from in-focus and out-of-focus configurations are shown in Figure\,\ref{fig:exp_setup}. Eye trackers for XR devices use similar cameras, and this form of eye tracker is readily available to researchers and consumers. This is one instance of a secure eye tracking configuration. An example alternative configuration would be using an eye camera with an adjustable focus lens. 

\textbf{Camera Distance}
The out-of-focus configuration was implemented by increasing distance between the eye and camera to degrade iris authentication. First, the in-focus configuration was set up by placing the eye camera as close as possible to the user's eye, while keeping the eye in the center of the eye image frame. Then, to create the out-of-focus configuration the experimenter adjusted the telescoping arm to the farthest point, again orienting the camera such that the eye stayed within the frame. We compute the distance between the camera lens and the eye to quantify the impact of this process on gaze accuracy and iris authentication.

\begin{figure}[!ht]
    \centering
    \includegraphics[width=\linewidth]{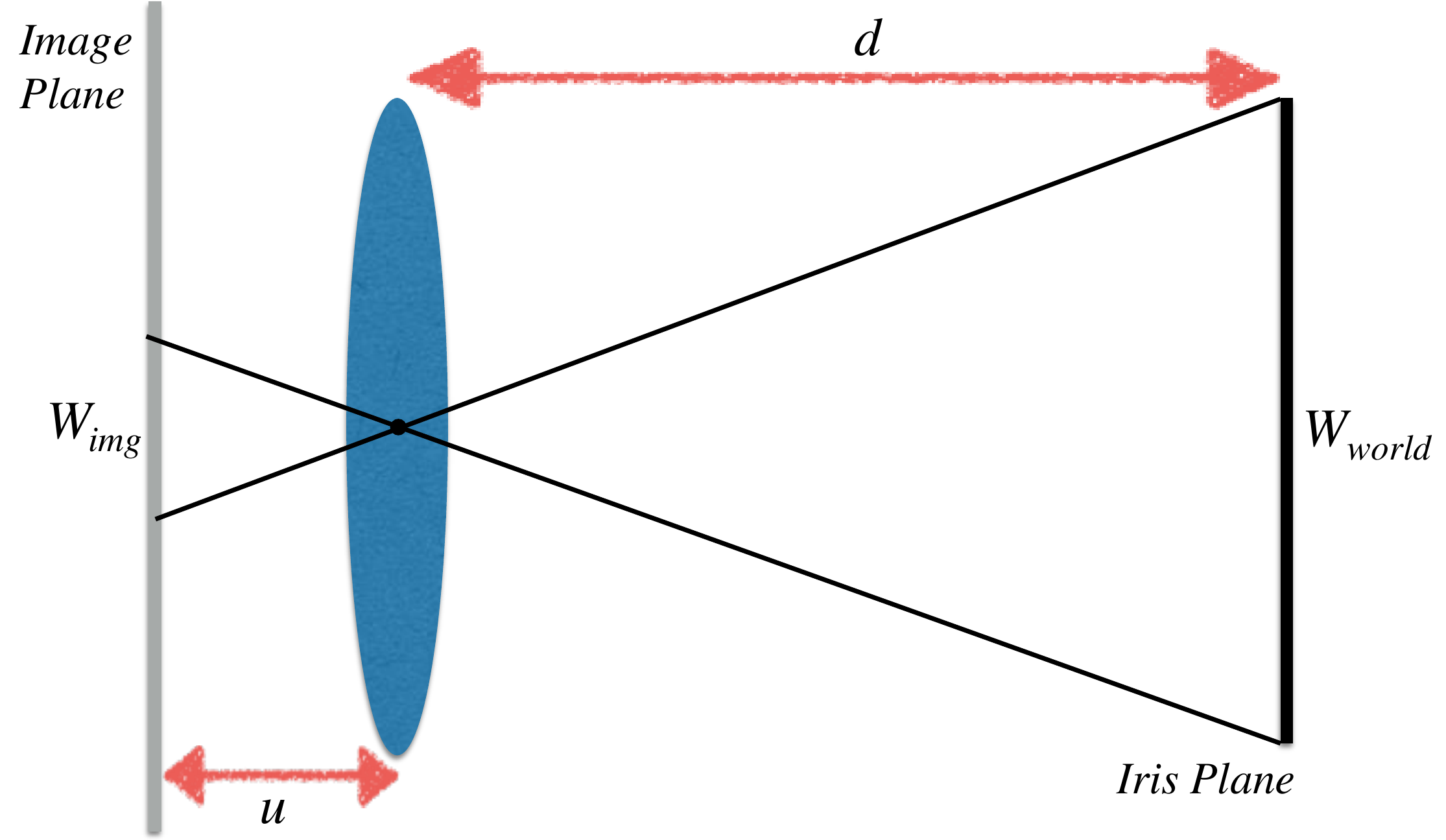}
    \caption{Camera system imaging a flat iris plane with a thin lens. The lens equation produces the relationship $\frac{1}{f}$=$\frac{1}{d}+\frac{1}{u}$ and $\frac{W_{img}}{u}$=$\frac{W_{world}}{d}$.}
    \label{fig:camera_distance}
 
    
\end{figure}

Camera distance is computed by modeling an imaging system with a thin lens. Figure\,\ref{fig:camera_distance} illustrates such a system. The distance between the iris and lens, and the lens and camera sensor are related by \begin{equation} \label{eq:two}
    \frac{W_{img}}{u}=\frac{W_{world}}{d},
\end{equation} where $d$ is the distance from the lens to the iris plane, $W_{img}$ is the width of the iris as measured in the image, and $W_{world}$ is the actual width of the iris. Variables $d$ and $u$ are related by the lens equation\begin{equation}
    \frac{1}{f}=\frac{1}{d}+\frac{1}{u},\label{eq:lens_equation}
\end{equation} where $f$ is the focal length of the camera in mm. We estimate $W_{world}$ as the average width of a human iris, $11$ mm\cite{nishino2006_journal}, measure $W_{img}$ within the image, and compute $f$. 

As shown in Figure\,\ref{fig:camera_distance}, the distance between the lens and camera sensor, $u$, is constant. The same process was followed to set up the in-focus eye tracker configuration for each participant. We compute $u$ as follows: (1) For each participant Eq.\,\ref{eq:two} and Eq.\,\ref{eq:lens_equation} are used to solve for $d$,\,(2) the average distance for all participants, $\overline{d}$, is computed,\,(3) $\overline{d}$ is substituted into Eq.\,\ref{eq:lens_equation} to compute $u$. This process assumes that the measured values of $d$ are within the depth of field of the camera for which the iris region is in-focus, and can be be estimated with the average distance, $\overline{d}$.

We computed the amount of defocus, $\sigma$, that was generated by an increase in distance from the in-focus configuration to the out-of-focus configuration. First, the out-of-focus distance $d_{secure}$ is computed using $\frac{W_{img}}{u}$=$\frac{W_{world}}{d_{secure}}$ and $W_{img}$ from an out-of-focus eye image. This is then substituted into Eq.\,\ref{eq:lens_equation} to generate a new $u$ value, $u'$, that represents the depth of a focal plane given the new camera distance. The amount of defocus $\sigma$ in mm is then computed by \begin{equation} \label{eq:sigma}
    \frac{u'}{D}=\frac{u-u'}{\sigma},
\end{equation}where $D$ is the lens diameter measured to be $1.05$mm, and $\sigma$ represents the spread of a point that was in focus at the near distance, projected onto the camera sensor at the distance $d_{secure}$. $\sigma$ is then converted from mm to pixels using the factor $0.003\frac{mm}{pixel}$, as specified in the OV9712-1D sensor spec sheet.

\textbf{Iris Authentication}
For our experiment, iris segmentation was performed using an open source implementation of IrisSeg\cite{gangwar2016irisseg}. Our authentication procedure applies a bank of 1D Log-Gabor filters to the resulting iris pattern to generate a binary code that captures the identifying features of the iris pattern\cite{masek2013matlab,kahlil2010generation}.

To perform authentication the bit values of these codes are compared using Hamming Distance to determine if the source and target match. Hamming distance is defined as the number of bits that disagree between source and target binary codes, \begin{equation}
\tvar{HD}=
\frac{\lVert (\tvar{S_{code}}\otimes\tvar{T_{code}})\cap(\tvar{S_{mask}}\cap\tvar{T_{mask}})\rVert}
     {\lVert\tvar{S_{mask}}\cap\tvar{T_{mask}}\rVert},
 \end{equation} 
where $S_{code}$ and $T_{code}$ are the input binary codes with their respective masks. The binary masks indicate which pixels contain the iris pattern, with zeros indicating eye lids, eye lashes, or any other detected noise\cite{daugman2009iris}. In the subsequent data analysis iris codes are excluded if at least 75\% of the bits are considered noise. We use a threshold of $HD_{auth}=0.37$ to authenticate a match between source and target.

    


\begin{table*}[!ht]
\centering
\begin{tabular}{|l|c|c|c|c|c|c|c|c|c|}
\hline
\
ID &   \begin{tabular}{@{}c@{}}\textbf{In-Focus} \\ \textbf{Distance}\\ \textbf{(mm)} \end{tabular}
& \begin{tabular}{@{}c@{}}\textbf{Out-of-focus} \\ \textbf{Distance}\\ \textbf{(mm)}\end{tabular} & \begin{tabular}{@{}c@{}}\textbf{Defocus}  \\ \textbf{$\sigma$}\\ \textbf{(pixels)}\end{tabular} &
\begin{tabular}{@{}c@{}}\textbf{In-Focus} \\ \textbf{CRR}\\ \textbf{(\%)}\end{tabular} &
\begin{tabular}{@{}c@{}}\textbf{Out-of-focus} \\ \textbf{CRR}\\ \textbf{(\%)}\end{tabular} & \begin{tabular}{@{}c@{}}\textbf{In-Focus} \\ \textbf{Gaze Error}\\ \textbf{($^\circ$)}\end{tabular} &     \begin{tabular}{@{}c@{}}\textbf{Out-of-focus} \\ \textbf{Gaze Error}\\ \textbf{($^\circ$)}\end{tabular} &
\begin{tabular}{@{}c@{}}\textbf{In-Focus} \\ \textbf{Precision}\\ \textbf{($^\circ$)}\end{tabular} &\begin{tabular}{@{}c@{}}\textbf{Out-of-focus} \\ \textbf{Precision}\\ \textbf{($^\circ$)}\end{tabular} \\ \hlineB{2.5}
S01      & 24.9 & 32.4 & 2.5 & 91 & 16 & 1.0 & 1.0 & 0.1 & 0.1 \\ \hline
S02      & 30.0 & 35.8 & 3.6 & 87 & 3  & 1.4 & 1.3 & 0.1 & 0.1 \\ \hline
S03      & 24.3 & 32.6 & 2.5 & 90 & 5  & 1.2 & 2.4 & 0.1 & 0.1 \\ \hline
S04      & 24.6 & 31.7 & 2.2 & 95 & 15 & 2.2 & 2.0 & 0.1 & 0.1 \\ \hline
S05      & 34.3 & 37.1 & 3.9 & 42 & 0  & 1.5 & 1.8 & 0.2 & 0.1 \\ \hline
S06      & 26.1 & 38.8 & 4.4 & 73 & 0  & 1.7 & 1.1 & 0.1 & 0.1 \\ \hline
S07      & 26.3 & 33.4 & 2.8 & 82 & 21 & 1.4 & 2.7 & 0.1 & 0.1 \\ \hline
S08      & 29.4 & 36.3 & 3.7 & 78 & 1  & 1.4 & 1.6 & 0.1 & 0.1 \\ \hline
S09      & 24.6 & 37.6 & 4.1 & 66 & 1  & 1.8 & 1.4 & 0.2 & 0.2 \\ \hline
S10      & 24.0 & 34.7 & 3.2 & 86 & 0  & 1.1 & 1.4 & 0.1 & 0.1 \\ \hline
S11      & 27.5 & 37.8 & 4.1 &  99 & 0  & 0.8 & 1.5 & 0.1 & 0.2 \\ \hline
S12      & 25.4 & 33.7 &  2.9 & 78 & 27 & 2.0 & 2.1 & 0.1 & 0.1 \\ \hline
S13      & 26.5 & 33.9 &  3.0 & 76 & 17 & 1.9 & 1.5 & 0.1 & 0.2 \\ \hline
S14      & 24.9 & 37.4 &  4.0 & 79  & 0  & 0.9 & 1.6 & 0.1 & 0.1 \\ \hline
S15      & 30.0 & 31.8 &  2.2& 57 & 0  & 0.9 & 1.9 & 0.1 & 0.1 \\ \hlineB{2.5}
\textbf{Mean}   & 25.1 & 33.1 & 3.3 &  79 & 7  & 1.4 & 1.7 & 0.1 & 0.1 \\ \hlineB{2.5}
\textbf{Std.\,Dev.} & 2.8  & 2.3  & 0.7 &  15 & 9  & 0.4 & 0.5 & 0.04 & 0.04 \\ \hline
\end{tabular}
\caption{Security and utility results for in-focus and out-of-focus eye tracking configurations. On average there was a difference of $8$mm between in-focus and out-of-focus configurations. Defocusing the camera caused a decrease in CRR without an appreciable impact on gaze accuracy.}
\label{tab:hardware_iris_auth}
\end{table*}
\begin{figure*}[!ht]
    \centering
    \includegraphics[width=0.33\linewidth]{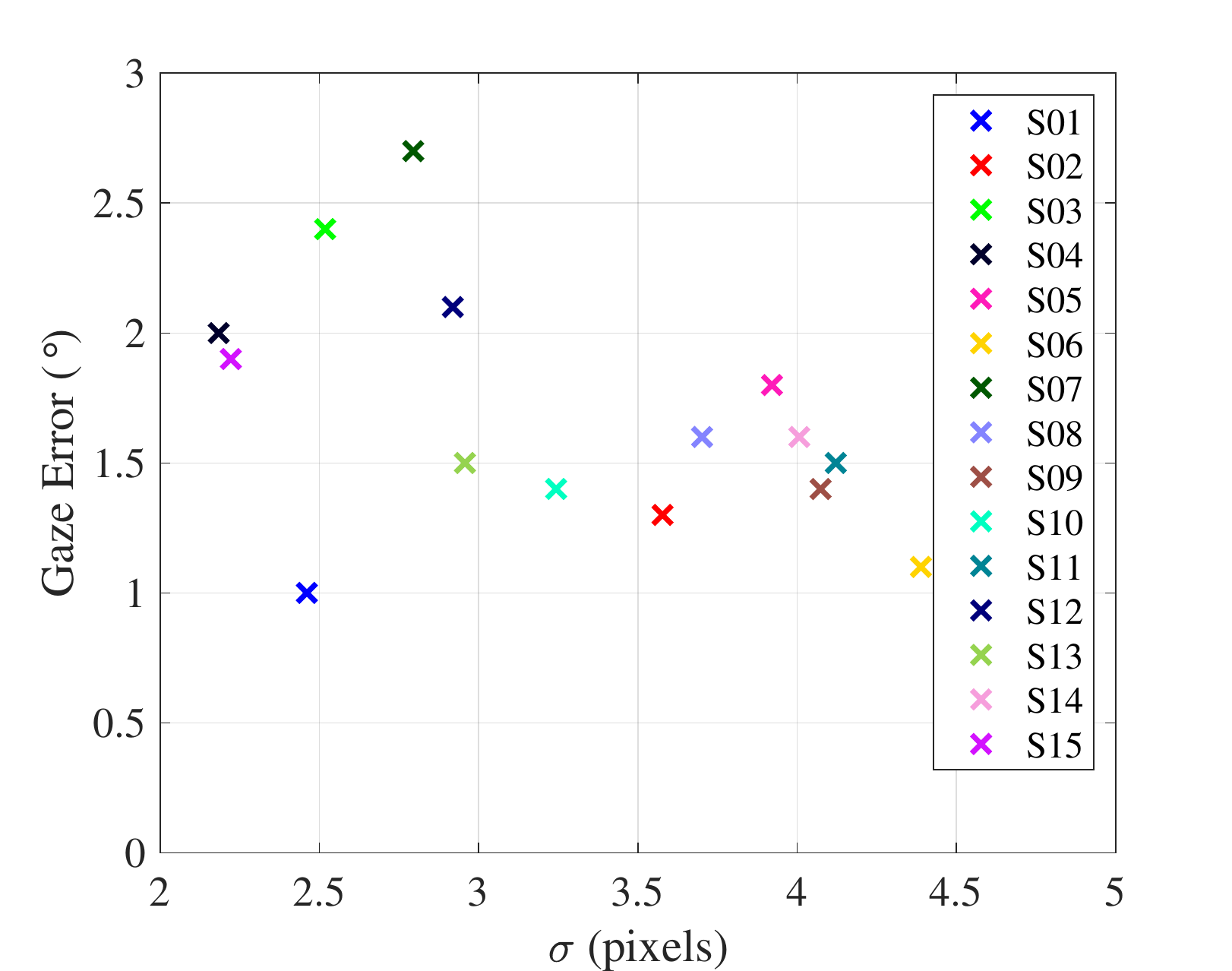}\includegraphics[width=0.33\linewidth]{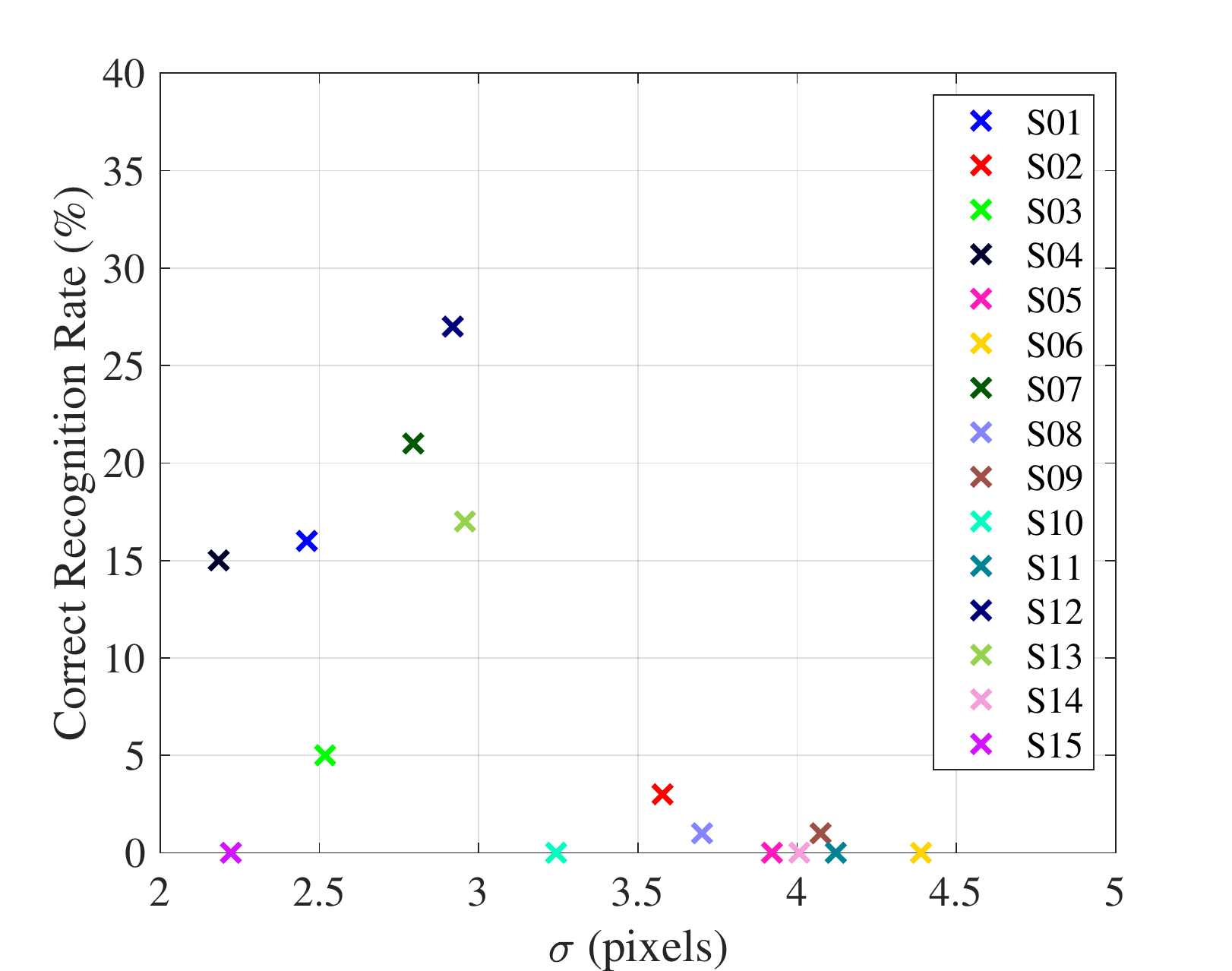}\includegraphics[width=0.33\linewidth]{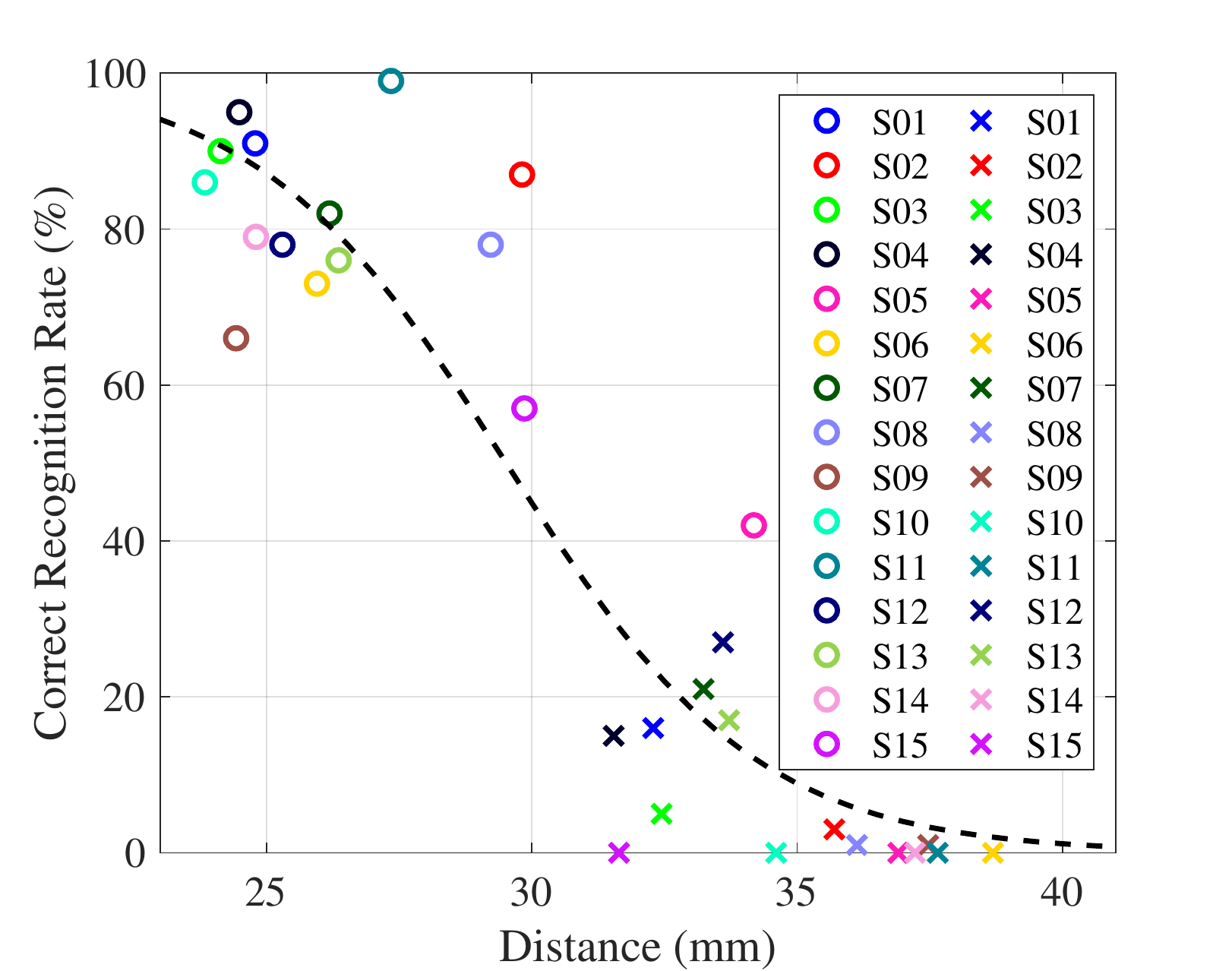}
    \caption{Results for security and utility show that CRR is degraded by defocus\,($\sigma$) and increased camera distance. Circles indicate data from the in-focus configuration, while crosses indicate data from the out-of-focus configuration. The dashed line represents a sigmoid curve fit to CRR as a function of distance. Angular error measured between targets and gaze data for the out-of-focus configuration was at most $2.7^\circ$.}
    \label{fig:auth_distance} 
\end{figure*}

\subsection{Evaluation}
\label{sec:secure_eval}
An iris authentication procedure is used to evaluate the increase in security from an in-focus configuration to out-of-focus configuration. Utility is measured using a target viewing task in a typical eye tracking setup, with error calculated between the estimated gaze positions and on-screen targets. Ideally, a secure configuration will degrade iris authentication while preserving the accuracy of gaze estimation.

\textbf{Metrics}
The ability to authenticate a user is measured using the Correct Recognition Rate\,(CRR)\cite{monro2007dct}. CRR is computed as the percentage of frames between the source and target inputs where $HD < HD_{auth}$. Through this metric we determine that images collected during our ``stop-and-stare'' authentication routine can be used to identify the individual, with minimal false positives. 

Eye tracking utility was measured in terms of gaze accuracy during the five target viewing task. The Pupil Labs software was used to identify frames with circular targets. These frames were used to calibrate a gaze mapping model that predicts the 2D gaze point-of-regard within the scene camera feed\cite{kassner2014pupil}. Once calibrated, the average error between projected gaze and the center of each target was computed in terms of visual angle. We computed the precision as the Root Mean Square Deviation between successive gaze locations while the targets were present to measure the stability of the gaze at each target.

\textbf{Method}
We modify the experimental setup of John et al.\cite{John2019} and evaluate the proposed hardware solution. 
Users sat with a chin rest and viewed circular targets presented on a desktop screen\,(see Figure\,\ref{fig:exp_setup}). First, the eye tracker was set up for an in-focus configuration, as specified in Section\,\ref{sec:auth_implement}. A video with five circular targets appearing for four seconds each was then shown, generated with Pupil Labs. The eye tracker was calibrated offline using Pupil Lab's default $3d\_calibration$ routine from Pupil Labs, with gaze samples and ground truth collected when the targets were present. 

The user was asked to look directly at the eye tracking camera for five seconds, simulating a ``stop-and-stare'' interface for iris authentication\cite{daugman2007new,proenca2009ubiris}, prior to target viewing and directly afterwards. Only images from this part of the data collection were used for authentication, ensuring that the pupil and iris are on-axis with the camera. On-axis images increase the reliability of iris segmentation and matching\cite{thompson2013off,daugman2007new}. Each user logged around 300 frames during this procedure. 

Eye tracking data was collected at 30 Hz using a Pupil Labs Pro glasses-based eye tracker\,(ca. 2016) with an eye image resolution of 320x240\cite{kassner2014pupil}. We calibrated the fixed focus Pupil Labs eye camera using a checkerboard pattern and MATLAB's Single Camera Calibrator App to compute a focal length, $f$, of $338.04$ pixels\,($1.014$mm). Prior to analysis, frames containing blinks or motion blur were removed.

\begin{figure}[!ht]
    \centering
    \includegraphics[width=\linewidth]{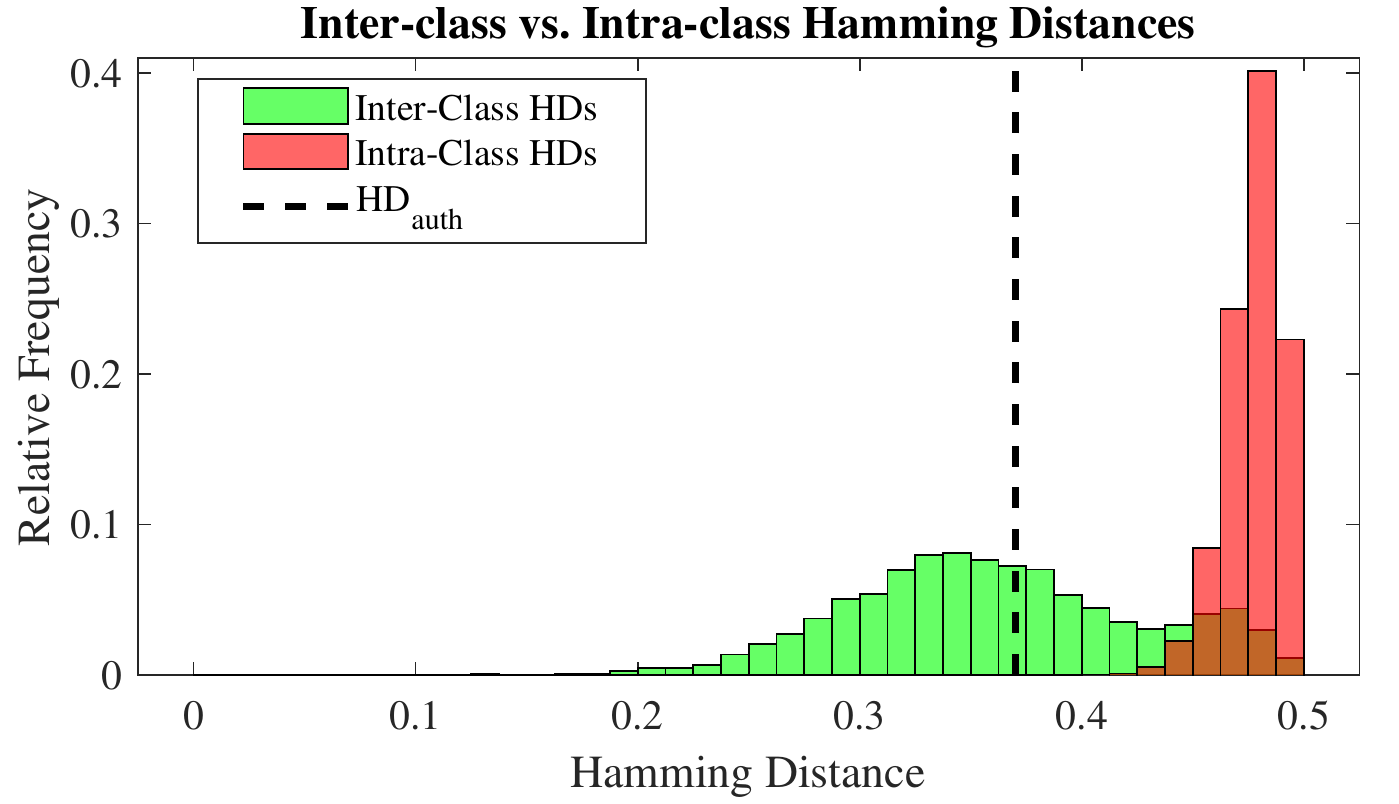}
    \caption{A histogram of inter-class and intra-class Hamming Distances is used to determine a threshold for iris authentication.} 
    \label{fig:hd_distributions}
    \centering
    \includegraphics[width=\linewidth]{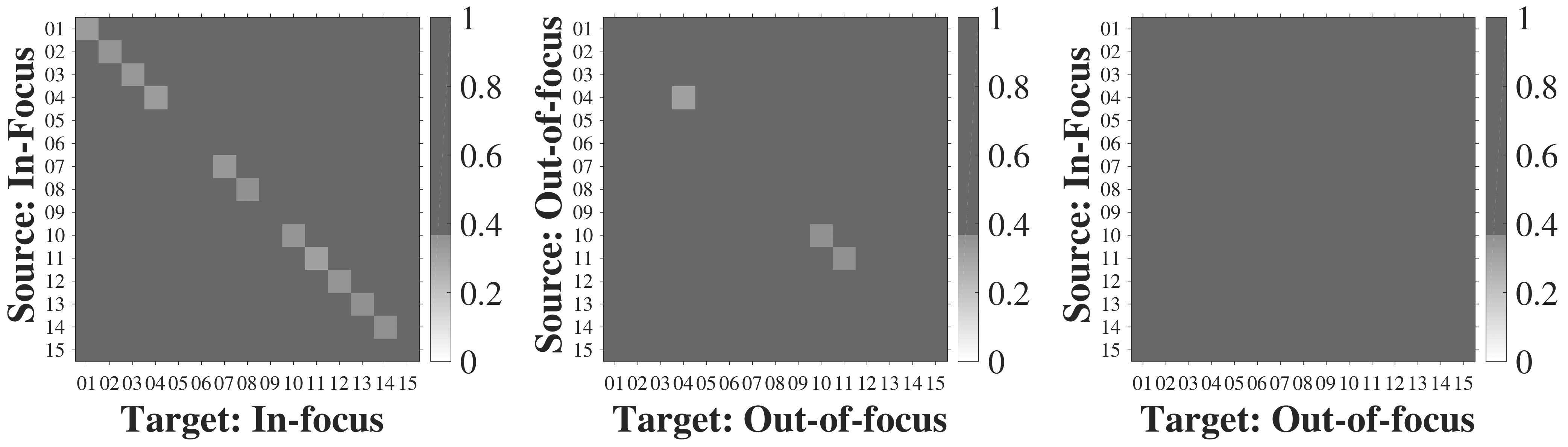}
    \caption{Average Hamming Distance between source and target comparisons between each individual. Values less than the $HD_{auth}$ indicate a match, and are colored white.} 
    \label{fig:authentication_matrix}
\end{figure}

\textbf{Participants}
Eye tracking data and images were collected from fifteen participants\,($8$ male, $7$ female) in an IRB approved user study. Participant demographics were $20$\% Asian, $13$\% Hispanic, $13$\% African American, $27$\% Indian, and $27$\% Caucasian.

\textbf{Results}
We computed an authentication threshold, $HD_{auth}=0.37$, based on the distributions of inter-class and intra-class $HD$ values for our in-focus eye images. These distributions and $HD_{auth}$ are illustrated in Figure\,\ref{fig:hd_distributions}. For our entire dataset, $HD_{auth}=0.37$ creates an overall true positive rate of 60.1\%, a false negative rate of 39.9\%, true negative rate of 99.9988\%, and a false positive rate of 0.0012\%. While a higher true positive rate may be ideal, increasing the value of $HD_{auth}$ would also increase the the false positive rate and compromise the system.

The average Hamming Distance between participants $i$ and $j$, $\overline {HD_{ij}}$, is shown in Figure\,\ref{fig:authentication_matrix}, with white cells indicating $\overline {HD_{ij}}$ is less than $HD_{auth}$. Grey cells indicate that the source does not match the target. Our results generate no false positives when the authentication condition is $\overline {HD_{ij}} < HD_{auth}$. 

Authentication by $\overline {HD_{ij}}$ has the highest accuracy when comparing in-focus images with in-focus images. As expected, in-focus images did not create any matches with out-of-focus images. However, out-of-focus images did create matches with other out-of-focus images from the same individual, albeit less frequently. Only three participants produced a $\overline {HD_{ij}}$ less than $HD_{auth}$ for the out-of-focus images.

Table\,\ref{tab:hardware_iris_auth} reports the CRR values, with an average in-focus CRR of 78.6\%, while the out-of-focus images had a rate of 7.1\%. Figure\,\ref{fig:auth_distance}\,(Right) demonstrates the relationship between camera distance and CRR by fitting a sigmoid function of the form $f(d) = \frac{1}{1+e^{-(a\cdot d + b)}}$, where $a=-0.43$,  $b=12.10$, and $d$ is the input distance in mm. At $30$mm CRR was $45$\%, and by $35$mm CRR has dropped to $8$\%, showing that only a small percentage of frames can successfully authenticate the user at increased distances.

The computed $\sigma$ values from Eq.\,\ref{eq:sigma} are presented in Table\,\ref{tab:hardware_iris_auth}. Figure\,\ref{fig:auth_distance}\,(Left\,\&\,Center) shows gaze accuracy and CRR respectively for the out-of-focus configuration as function of $\sigma$. Using $\sigma$ to measure defocus allows us to compare results across configurations independent of the implementation, such as with software-based Gaussian blur.

Table\,\ref{tab:hardware_iris_auth} contains these average gaze error and precision values for each participant and configuration and Figure\,\ref{fig:auth_distance}\,(Left) demonstrates the relationship between out-of-focus gaze error and $\sigma$. We found that the average error across participants for the in-focus and out-of-focus configurations were 1.4$^\circ$ and 1.7$^\circ$ respectively. These values both fall within the magnitude for noise in an eye tracking system\cite{holmqvist2012eye}. Ten participants saw an increase in error from the out-of-focus configuration, with the maximum error being 2.7$^\circ$. Error within this range is acceptable for target viewing, where the targets span approximately 7$^\circ$.

\textbf{Discussion}
Using $HD_{auth}$ we computed CRR for each participant, comparing every eye image with every other eye image. For the in-focus configuration we found on average 78.6\% of frames were a match. This indicates that a login procedure that matches only one input image to another may not be robust enough for a consistent user experience. Using a larger $HD_{auth}$ would create a smoother process, but compromise security. Instead, collecting a small set of ideal on-axis images and computing the average hamming distance from the reference may be a more dependable approach. 


Previous research has shown that camera distance degrades iris authentication\cite{John2019}, based only on data from one user. We paramaterized and evaluated this approach extensively, and found that the average error of 1.7$^\circ$ introduced with this secure configuration did not have considerable impact on a target viewing task. While this error is acceptable for target viewing, it may not be for more complex applications such as animating the eyes of a social virtual avatar.
\section{Evaluation of Utility for Avatar Gaze}
Social virtual avatars have eye animation with the goal of increasing social presence and immersion. A large body of work has established that the animation of eye movements impacts viewer perceptions of avatar attributes, such as truthfulness and attentiveness\cite{ruhland2015review,ferstl2017facial,steptoe2010lie}. The goal of this section is to determine how the noise introduced by secure eye tracking impacts the perception of animated virtual avatars.

\begin{table*}
\centering
\begin{tabular}{|l|c|p{14cm}|}
\hline
\textbf{ID} & \textbf{Video Timestamp} & \textbf{Sentences spoken} \\ \hlineB{2.5}
1            &  0:30.73 - 0:42.73&  {\small  ``You may have been wondering, well, how am I going to learn to speak English properly if I don't study English grammar? Well, today I am going to talk to you about that. So, how is it that we do this?"  }            \\ \hline
2            & 0:44.26 - 0:55.26   & {\small ``There is a special technique and it is very easy. Research has shown that it is the best way to learn English grammar, or grammar for any language."   }             \\ \hline
3            & 0:59.51 - 1:11.51 & {\small``It's called point of view stories, or point of view mini stories. A mini story is just a small story. So, how is that we use point of view stories to learn English grammar?"} \\ \hline
4            & 1:20.10 - 1:32.10 & {\small``We listen to a number of different points of view for this very story. And by point of view I mean that we change something in the story, like the time that the story is being told."} \\ \hline
5            & 1:47.58 - 1:59.58 & {\small``Let's start with an example. Now, when I usually teach these in a classroom, I'll start by telling it in the present tense. So let's start there."}\\ \hline
6            & 2:28.18 - 2:40.18 & {\small``Make sure you understand it. So your next question may be, how are we going to use this story to learn English grammar? And that's a good question. As I said before, we are going to hear this story told in a number of different ways."} \\ \hline
\end{tabular}
\caption{Conversational sentences spoken and their timestamps in the video\textsuperscript{\ref{ft:youtube}} while eye movements were recorded for each animation stimulus.}
\label{tab:stimuligeneration}
\end{table*}

\subsection{Research Questions \& Expected Outcomes}
We conducted perceptual studies to answer the following questions:
\begin{itemize}
    \item $RQ_1$: At what level of defocus do viewers detect a difference in the animation of a virtual avatar's eyes compared to a reference?
    \item $RQ_2$: What is the relationship between eye image defocus and the perception of avatar truthfulness, naturalness, attentiveness, comfort, and eye contact?
\end{itemize}
For $RQ_1$ we hypothesize that a medium amount of defocus, i.e., less than or equal to $\sigma = 3$ pixels, applied to the image feed will be detected by the viewer at a rate near chance, while data from larger values of $\sigma$ will be detected at a higher rate. Past work has shown that the pupil detection rate declines after $\sigma=3$\cite{John2019}, which would result in a halt in eye animations during frames where the pupil was not detected. Additionally, the offset in gaze required for a viewer to indicate there is no longer mutual gaze varies across viewing distance and display mediums, ranging from less than $1^\circ$ up to $9^\circ$\cite{gamer2007you}. Our results in Section\,\ref{sec:secure_eval} indicate gaze error for all values of $\sigma$ up to $4.4$ were more than $1^\circ$ and less than $3^\circ$, falling within the range for mutual gaze with a virtual human face. 

For $RQ_2$ we selected the attributes truthfulness, naturalness, attentiveness, comfort, and eye contact as they are influenced by avatar eye movements. We again hypothesize that up to a medium amount of defocus, i.e., less than or equal to $\sigma = 3$ pixels, there will be no difference in how eye movements are perceived by viewers. For defocus greater than $\sigma=3$ we expect negative responses, or values less than $3$\,(`Neither Disagree or Agree') on the measured Likert scale.



\subsection{Study 1: Detection Threshold}
\label{sec:study1}

The goal of study 1 is to answer $RQ_1$. We designed a same-different experiment where naive viewers are presented with a reference avatar with unmodified eye tracking and a stimulus avatar with modified eye tracking, and they are tasked with reporting whether the two avatars are identical or different from each other. We compute psychometric curves from the participant responses and report the point of subjective equality\,(PSE) and detection threshold\,(DT). These values clarify the level of defocus at which viewers are able to perceive a difference in the eyes of the virtual avatar.

\textbf{Stimuli Generation}
\label{sec:study1_stimuli}
Naturalistic gaze data was recorded with the Pupil Labs Pro glasses-based eye tracker in a conversational scenario. We selected an English as a second language instructional video from YouTube\footnote{\label{ft:youtube}\url{https://tinyurl.com/yxetvjw8}}. The details of the video are shown in Table\,\ref{tab:stimuligeneration}. The video had an instructor speak conversational sentences in English for the student to pause and repeat back to them. The topic of the conversation was a technique for learning English grammar. One of the authors watched the video, and acted out the part of the student by repeating sentences back as appropriate while being eye tracked. We extracted six 12 second segments from different parts of this dataset, resulting in six eye animations. Gaze directions from these segments were transferred on to a virtual avatar.


The virtual avatar was animated and rendered with the Unity game engine, version 2017.4.24f1. We created a model with Character Creator 3. We chose a bald male avatar with a realistic appearance to avoid simulating hair. Only the eyes of the avatar were animated; the rest of the face was static without any eye movement. The model was rigged to animate both eyes using monocular gaze data, as our eye tracker only records movements of the right eye. A reference gaze vector, $<x,y,z>$, was recorded with the author looking straight ahead at the beginning of data collection. This vector is used to generate a gaze offset vector, $<-x,-y,0>$, which when added to the reference gaze direction creates a `forward' vector $<0,0,z>$. Using the `forward' vector the avatar eye is oriented straight ahead towards the viewer. For each gaze vector in the animation data stream the gaze offset is added to generate a current gaze direction, which is then used to orient each eye. Gaze shifts relative to the gaze offset create the eye movements seen in the stimuli. The animations did not include any audio.

\begin{figure}[ht]
    \centering
    \includegraphics[width=\linewidth]{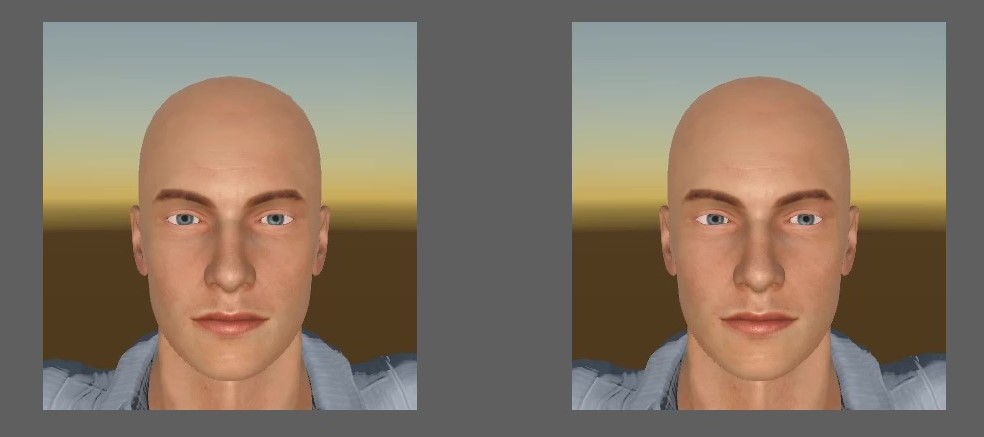}
    \caption{Eye animations from blurred and unblurred images were presented side-by-side with identical avatars. The deviation in gaze from defocus\,($\sigma=5$) is shown in the avatar on the right.}
    \label{fig:same-different-interface}
\end{figure}

For each of the six eye animations, we had the recorded eye images from which gaze positions are estimated. The original eye images were blurred using MATLAB's $imgaussfilt$ function at five levels of $\sigma$, defined in pixels. The blurred images were fed through the Pupil Pro gaze detection pipeline, and the gaze positions estimated from the modified eye images were recorded. This procedure ensured that the internal parameters in the processing pipeline for the eye images were constant. It may be possible to tune the internal parameters in the processing pipeline to fit blurred input eye images. We left this tuning and its evaluation for a future experiment.

The Unity camera was positioned such that the rendered face spanned 8 visual degrees in our experimental setup, consistent with the size of a human face at a distance of $1.5$m. This design choice was based on prior studies that use a distance of $1.5$m or more when evaluating real face to face conversations and virtual interactions\cite{bennett2016looking,macquarrie2019perception}. Figure\,\ref{fig:same-different-interface} shows the experiment setup with the side by side avatars. The reference and stimulus were not labeled. The left and right placement was swapped to avoid bias, resulting in a total of 60 trials per participant\,($6$ eye animations $\times$ $2$ positions\,(L/R) $\times$ $5$ levels of $\sigma$).

\textbf{Method}
We created a same-different task where each trial consists of a stimulus and a reference presented simultaneously, and participants are asked to indicate if they are the same or different\cite{duchowski2015eye,cheetham2011human}. The Miss Rate is computed as the proportion of `same' responses. The Point of Subjective Equality\,(PSE) corresponds to the 50\% Miss Rate, i.e., the stimulus level at which participants are as likely to detect as they are to miss the difference between the stimulus and the reference. In other words, the PSE clarifies when a viewer can discriminate the presence of the stimuli at the same rate as chance. The Detection Threshold\,(DT) provides an upper bound on the amount of defocus that can be applied before a difference is perceived by a viewer.

We select the value of $\sigma$ that corresponds to a 25\% Miss Rate as the DT, where the participant is expected to consistently respond that the two animations are different. This threshold has been previously adopted in several virtual reality pyschometric experiments, as it is halfway between chance and a perfect detection rate\cite{bolte2015subliminal,steinicke2009estimation,bolling2019shrinking,zenner2019estimating}.

Our experiment uses a within-subjects design. Participants were presented with avatar eye animation generated from five levels of defocus in a randomized order. Both the stimuli and reference were shown on screen at the same time. Before starting the experiment participants were given the following prompt to describe the task: ``In each trial you will be shown two videos of an animated virtual avatar side by side for 12 seconds. Only the eyes will be animated. Your task is to indicate whether the two animations presented are the same or different. Again, only the eyes are animated, so there will be no differences in other regions of the face. You will provide your response after each trial.'' A break was offered to participants halfway through the experiment. After completing all of the trials, participants filled out a post-study questionnaire that indicated their age, gender, ethnicity, and prior experience with virtual reality displays. The experiment took approximately 25 minutes. The experimental setup was as shown in Figure\,\ref{fig:exp_setup}. Eye tracking data from a remote device mounted to the monitor was recorded, but is not discussed in our analysis.


\begin{figure}[!ht]
    \centering
    \includegraphics[width=\linewidth]{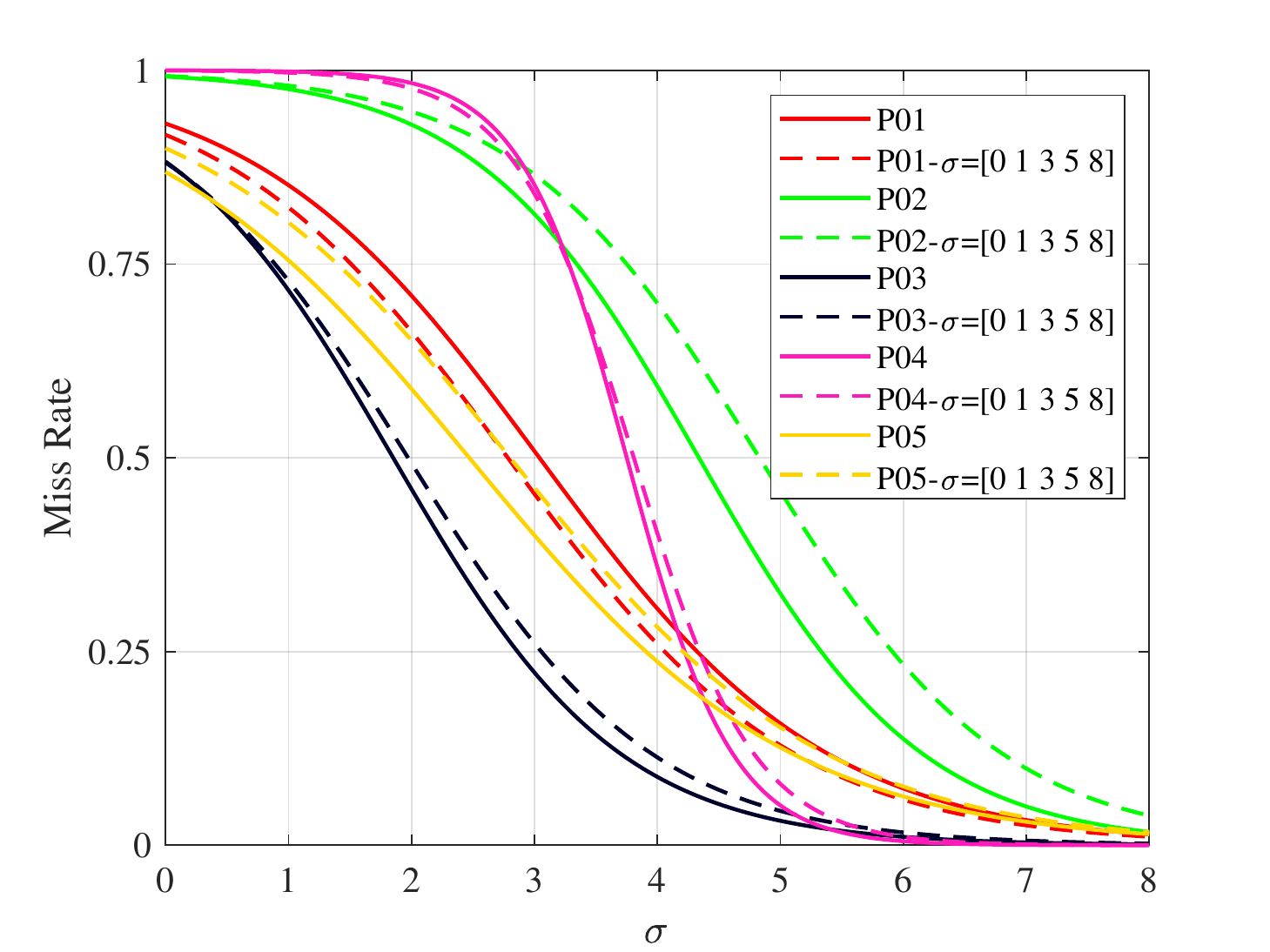}
    \caption{Study 1 pilot results. Colored solid lines indicate psychometric functions fit to responses for all defocus levels, $\sigma=[0,1,2,3,4,5,6,7,8]$, for five participants and dashed lines indicate the corresponding functions fit to the subset $\sigma=[0,1,3,5,8]$.}
    \label{fig:pilot_study1}
\end{figure}

\textbf{Pilot experiment} A five participant pilot experiment was conducted to establish the number of defocus levels ($\sigma$) to evaluate in Study 1. Nine levels, $\sigma=[0,1,2,3,4,5,6,7,8]$, were used. We found that the experimental duration exceeded 30 minutes and resulted in participants reporting fatigue even with mandatory breaks. Individual psychometric functions are shown in Figure\,\ref{fig:pilot_study1}. We then computed individual psychometric functions if only a subset of these nine levels were presented to the participant. We found that using only the defocus levels corresponding to $\sigma=[0,1,3,5,8]$ led to psychometric curves that were comparable\,(varied by at most 0.6 $\sigma$). Reducing the number of defocus levels reduced the the experiment duration by approximately 11 minutes. Hence the subset was selected for the main experiment. 

\begin{figure}[ht]
    \centering
    \includegraphics[width=\linewidth]{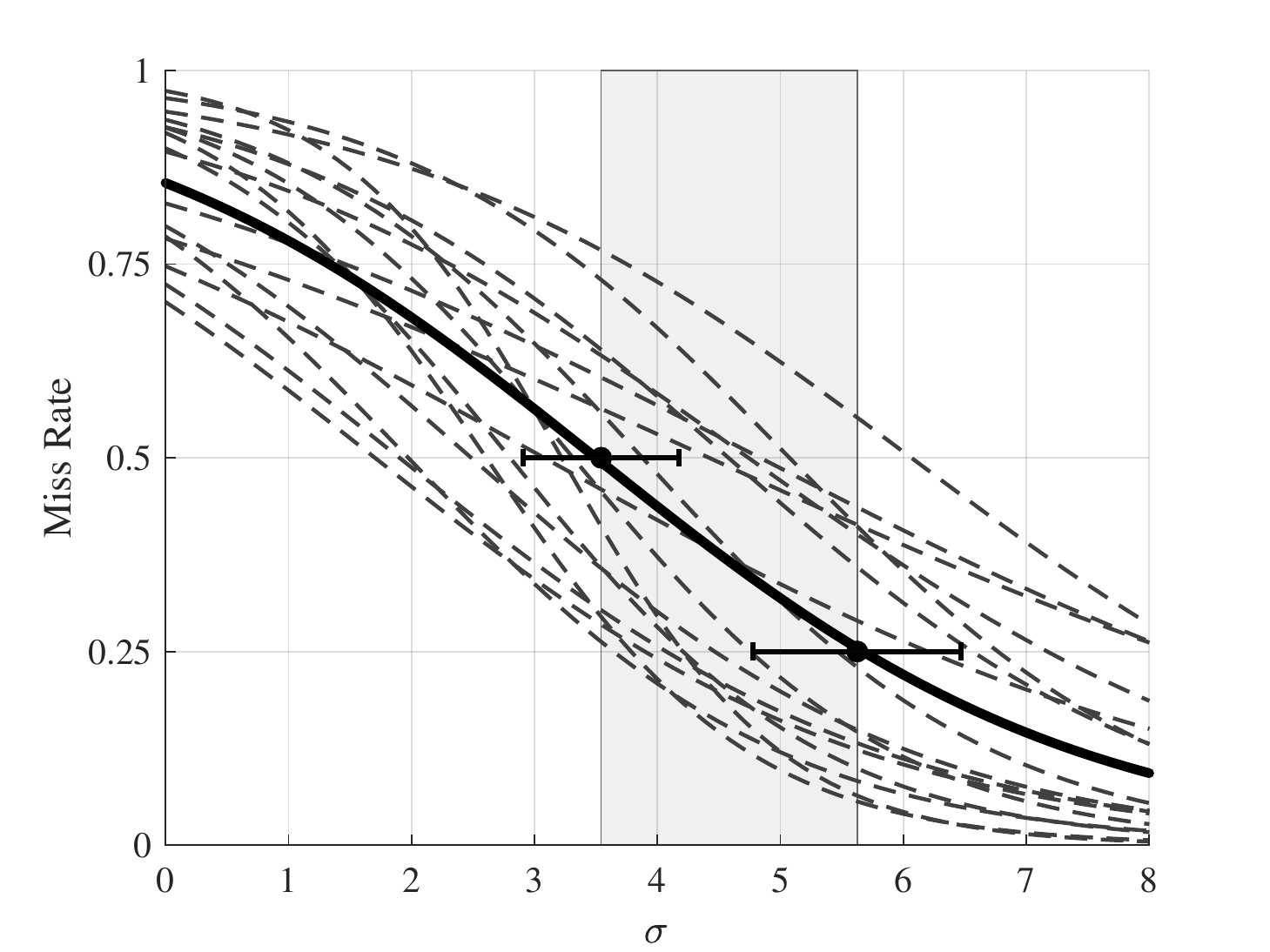}
    \caption{Resulting psychometric functions of the same-different task for individual and pooled responses. Gray dashed lines represent individual responses, and the solid black line represents a function fit to the average responses across individuals. Error bars represent the 95\% confidence interval for PSE and DT values.}
    \label{fig:psychometric_func}
\end{figure}

\textbf{Participants}
Twenty participants\,($11$ male, $8$ female, $1$ ``Preferred Not To Answer'') with age ranging from 18 to 26 years were recruited from the university community under an IRB approved protocol. All participants reported normal or corrected-to-normal vision. Four participants reported no prior experience with a VR HMD. Of the participants that had experience with VR HMDs, the majority\,(75\%) have used more than one type of HMD, such as the Oculus Rift DK2, Oculus Go, Google Cardboard, Samsung Gear VR, and HTC Vive.


\textbf{Results}
\label{sec:study1_results}
The responses collected from each participant consisted of 60 categorical data items\,(`same'/`different'). Responses for each participant were grouped by level of defocus. Miss Rate was then calculated for each $\sigma$ as the number of `same' responses divided by the total number of responses. Miss Rate represents the probability that the participant does not detect a difference between animations. 

Response quality was validated using the Miss Rate for stimuli where $\sigma=0$. These stimuli showed two identical animations on screen, and if the participant responded `different' 50\% of the time or more, they either misunderstood the task or did not follow instructions. Four participants were removed from analysis using this criteria. A psychometric sigmoid function, defined as $f(\sigma) = \frac{1}{1+e^{-(a\sigma+b)}}$, models the probability that a participant would answer `same' as a continuous function of $\sigma$. We fit $a$ and $b$ to each participant's responses using MATLAB's $glmfit$ function. PSE and DT are then computed for each individual.  
A pooled psychometric function was computed by averaging Miss Rate across participants, and then fitting a function to the values, see Figure\,\ref{fig:psychometric_func}. 
The shaded region indicates a usable range of defocus for each individual curve, marking the area between the PSE and DT where a difference is not consistently perceived.







\begin{table}
\centering
\begin{tabular}{|l|c|c|c|l|c|c|c|} \hline
\textbf{Participant} & \textbf{PSE} & \textbf{DT} & \textbf{Participant} & \textbf{PSE} & \textbf{DT} \\ \hlineB{2.5}
S01 & 3.22 & 4.21 & S14 & 2.61 & 3.79 \\ \hline
S02 & 1.90 & 4.07 & S15 & 1.70 & 3.90 \\ \hline
S05 & 4.58 & 6.56 &S16 & 2.80 & 4.21 \\ \hline
S06 & 4.42 & 8.21 &S17 & 3.31 & 4.75 \\ \hline
S07 & 6.06 & 8.38 &S18 & 1.97 & 3.64\\ \hline
S08 & 4.74 & 7.17 &S19 & 4.84 & 8.21\\ \hline
S09 & 3.08 & 6.20 & & &  \\ \hline
S12 & 2.49 & 4.47 & \textbf{Average} & 3.50 & 5.67  \\ \hline
S13 & 3.88 & 5.46 & \textbf{Std. Dev.} & 1.30 & 1.73  \\ \hline
\end{tabular}
\caption{PSE and DT for each participant in Study 1.}
\label{tab:PSEs}
\end{table}

\begin{figure*}[!ht]
    \centering
    \includegraphics[width=\linewidth]{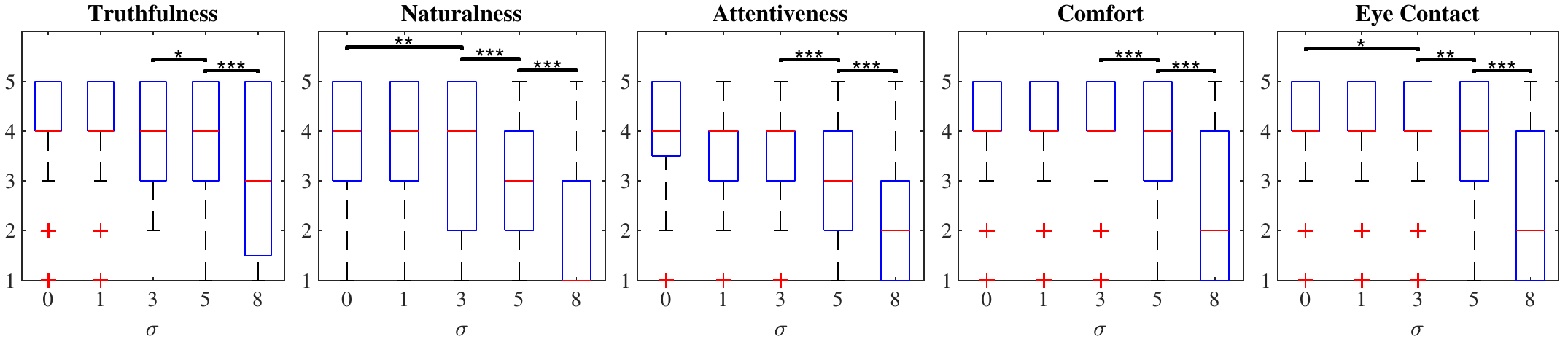}
    \caption{Box plots indicating the median, 25\%, and 75\% quartiles for Study 2 results. Significantly different groups are marked with \textbf{*} when $p<.05$, \textbf{**} when $p<.01$, and \textbf{***} when $p<.001$. For clarity ** significance brackets were established but not drawn between groups $\sigma$=$[0,1$] and $\sigma$=$5$, along with *** significance bars for groups $\sigma$=$[0,1,3]$ and $\sigma$=$8$ across all attributes.}
    \label{fig:study2_results}
\end{figure*}

\textbf{Discussion}
As shown in Table\,\ref{tab:PSEs}, the individual PSE values ranged from 1.70 to 6.06, and DT ranged from 3.64 to 8.38. The variation between individuals is also illustrated in Figure\,\ref{fig:psychometric_func}, as responses for Miss Rate ranges from 0.70 to 0.97 at $\sigma=0$, where there is no difference between the two presented animations. Another source of this variation might stem from the strategy participants used to detect differences. Several participants indicated after the experiment that they examined the amount of visible sclera on either side of the avatar's iris in both animations to determine if there was a difference, while others indicated they relied on the movement during shifts in gaze direction. We expect participant responses to be more accurate during our experiment than a typical interaction with an avatar, as they are informed to look for differences from a reference presented side by side with the stimuli.

Our analysis shows that the defocus value of $\sigma=3.50$ is the average PSE, i.e., at this defocus level the viewer has as much chance of perceiving difference in eye animation relative to the original reference as she does to not perceive any difference. The average DT is $\sigma=5.67$ pixels, which is the defocus level at which there is a 75\% chance that viewers will be able to detect that the eye animation of the avatar is different compared to the original. These findings connect well with the results reported by John et al.\cite{John2019}. They reported that defocus produced with $\sigma=3$  degraded iris authentication for most individuals and $\sigma=5$ completely degraded iris authentication. 

The implications of our findings taken together with John et al.'s reports are as follows: if a user is comfortable with a moderate level of security, they can use up to $\sigma=3.50$ of defocus without a noticeable effect on the eye animation of their social virtual avatar. Some users may want a higher level of security, and if they select a defocus of $\sigma=5$ there is some chance the noise will be noticed in exchange for their preferred security level. It is likely though that when the reference animation is not shown it will be more difficult for a viewer to notice that the eye animation is modified. It is possible that in this case viewers may feel that ``something is off'' and report that the avatar did not make eye contact with them, or that the avatar did not pay attention to them, or that the avatar was not truthful or natural. We investigated these judgements in our next experiment.


\subsection{Study 2: Avatar Attributes}
The goal of this experiment is to answer $RQ_2$. We measure responses to truthfulness, naturalness, attentiveness, comfort, and eye contact for increased values of $\sigma$. Five levels of $\sigma$ are considered: None\,($\sigma=0$), Low\,($\sigma=1$), Medium\,($\sigma=3$), High\,($\sigma=5$) and Very High\,($\sigma=8$).



\textbf{Stimuli Generation}
The animation renders from Study 1 were used for Study 2. However, instead of two animations being presented only one animation was shown at a time in the center of the screen.

\textbf{Method}
The study structure and apparatus was identical to Study 1, except participants rated each animation. The prompt provided to participants was: ``In each trial you will be shown a video of an animated virtual avatar for 12 seconds. Only the eyes are animated. Imagine you are having a conversation with the avatar. Your task is to respond to several prompts about the animation after each trial.''. Based on prior work we evaluate each interaction in terms of truthfulness\cite{steptoe2010lie}, naturalness\cite{jorg2018perceptual}, attentiveness\cite{ferstl2017facial}, comfort with the avatar\cite{guadagno2011social}, and eye contact\cite{mutlu2009footing}. After watching each animation the participant used a mouse to respond to the following prompts, using a five point Likert scale from `Strongly Disagree' to `Strongly Agree'($1$-$5$):

\emph{(1) The avatar was truthful. (2) The eye movements of the avatar were natural. (3) The avatar paid attention to me. (4) I felt comfortable in the presence of this avatar. (5) The avatar made eye contact with me.}

The experiment follows a within-subjects design, where every participant saw every animation and defocus level. Animations were presented in randomized order, and a break was offered halfway through the experiment. Each stimulus was presented and rated twice, leading to a total of 60 trials per participant\,($6$ eye animations $\times$ $2$ repetitions $\times$ $5$ levels of $\sigma$). The experiment took approximately 35 minutes. Again, eye tracking data from a remote device mounted to the monitor was recorded and is not discussed in our analysis.

\textbf{Pilot experiment} A five participant pilot experiment was conducted to establish the number of repetitions for Study 2. We found that showing three repetitions of each stimulus resulted in participants asking for additional breaks after the halfway point, and the experiment could take up to an hour to complete. To limit the experiment duration and reduce participant fatigue we decided to use only two repetitions.   

\textbf{Participants}
Nineteen participants\,($14$ male, $5$ female) with age ranging from 19 to 39 were recruited from the university community under an IRB approved protocol. All participants reported normal or corrected-to-normal vision. Participants were ineligible if they had previously participated in Study 1, to ensure they had not previously seen the animation stimuli.

\textbf{Results}
Likert scale responses for each dependent variable\,(truthfulness, naturalness, attentiveness, comfort, eye contact) represent ordinal data grouped by the defocus parameter $\sigma$. Figure\,\ref{fig:study2_results} shows the average and standard error values for each attribute.

The Kolmogorov-Smirnov test for normality was applied to each group and variable. Data were not normally distributed\,($p<0.001$), and therefore non-parametric statistical tests were used. A Friedman test showed a significant main effect of $\sigma$ for truthfulness\,($\chi^2(4)$=$162.72$,$p<0.001$), naturalness\,($\chi^2(4)$=$290.2$,$p<0.001$), attentiveness\,($\chi^2(4)$=$300.41$,$p<0.001$), comfort\,($\chi^2(4)$=$279.15$,$p<0.001$), and eye contact\,($\chi^2(4)$=$199.23$,$p<0.001$). For each attribute pairwise Wilcoxon signed rank tests with Bonferroni correction showed significant differences between $\sigma=5$ and all other levels of $\sigma$\,($p<.05$ or less); as well as between $\sigma=8$ and all other levels of $\sigma$\,($p<.001$). Additionally, for naturalness and eye contact significant differences were found between $\sigma=0$ and $\sigma=3$, with\,($p<.01$) and\,($p<.05$) respectively. Figure\,\ref{fig:study2_results} visualizes the results as boxplots.

\textbf{Discussion}
Our analysis shows that for $\sigma=0$, i.e., no blur, average responses for each attribute were approximately $4$, or `Slightly Agree' on the Likert scale. Thus, participants agreed that the avatar was truthful, eye movements were natural, the avatar paid attention to them, they were comfortable with the avatar, and maintained eye contact. They did not `Strongly Agree' with these statements, however. For truthfulness, naturalness, and comfort this is likely a result from only the avatar's eyes being animated and the lack of blinks. With respect to attentiveness and eye contact, the animations did not respond to the user's gaze, causing participants to provide only slight agreement. Still, at the end of the experiment several participants asked if the avatar was responding to their eye movements, as they knew they were being eye tracked. This indicates that the animation stimuli was convincing enough to simulate eye contact and interaction with the avatar.

Significance testing found a decrease in all response values at $\sigma=5$ and $\sigma=8$. Average responses for $\sigma=5$ ranged from $3.04$ to $3.78$, indicating participants did not have a negative experience, but less positive. This is consistent with the findings from Study 1, as $\sigma=5$ is near the DT\,($\sigma=5.67$). At $\sigma=8$ only responses for truthfulness averaged to $3$, i.e., `Neither disagree or agree', which means the participants were not able to consistently determine if the avatar acted truthfully based on eye movements. Avatars using $\sigma=8$ may be limited in their ability to immerse the viewer within a conversational setting. Averages for the rest of the attributes fell between $1.93$ and $2.58$, indicating that the avatar no longer convinced them. Eye tracking in the presence of this much defocus is not feasible for a convincing social avatar.
\section{Conclusion}
We have implemented and evaluated a novel hardware-based eye tracking configuration to secure the iris biometric from unauthorized identification. The secure configuration produced an average Correct Recognition Rate of $7$\% compared to $79$\% before defocus is introduced. Our second contribution is a pyschophysical experiment that determines the detection threshold for users viewing the eye movements of a virtual avatar animated using eye tracking data. Our results suggest that a defocus parameter of $\sigma=3.5$ should be used if utility is preferred over security, and $\sigma=5$ if security is preferred. Our third contribution is measuring the effect of $\sigma$ on several attributes important for social interactions with virtual avatars, such as eye contact and naturalness. Results indicate attributes are degraded at $\sigma$ values of $5$ and $8$, and the avatar no longer maintains eye contact, attentiveness, or naturalness.

\textbf{Limitations} The stimuli used for our perceptual evaluation has limitations. Particularly, our evaluation does not consider the impact of defocus on eye movement characteristics such as the blinks, the dynamics of saccades with large amplitudes, or estimated pupil diameter. These characteristics play an important role in complex social interactions and are more prominent the closer the user is to the avatar. The stimuli also did not include head or mouth movements. 
The defocus solution presented in this paper leveraged the telescoping arm of a popular eye tracker. More generally, a defocus solution applies to configurations where the eye camera is readily accessible, though future work might investigate clip on optics similar to Pittaluga and Koppal\,\cite{pittaluga2015privacy}. Our findings with respect to the fall in correct recognition rate are based on the Daugman method of iris recognition. If the iris recognition module were to be replaced with upcoming deep network based approaches, such as one proposed by Proenca and Neves\cite{proenca2019segmentation}, the fall in correct recognition rate as a function of hardware parameters might need to be re-assessed. Our work provides a foundation for developing an automated system that continually optimizes the security-utility trade-off even as new methods of eye tracking and iris recognition are invented.


\textbf{Future Work} It would be interesting to investigate an optimization framework for security and utility. It would also be useful to create implementations of secure eye tracking configurations that apply to different camera form factors. Additional perceptual experiments with a smaller distance from an avatar, and more realistic features on the avatar that include blinks, eyelid movements, and pupil diameter would provide further insight as well as implementing a similar evaluation within an immersive VR environment. Our work motivates active research in these directions \textit{before} eye tracking in XR becomes ubiquitous and users are at risk to malicious attacks.


\acknowledgments{The authors wish to thank Cody LaFlamme for generating the stimuli used in our studies. Authors acknowledge funding from the National Science Foundation\,(Awards IIS-1566481, IIS-1514154, and IIS-1423189), and the NSF Graduate Research Fellowship\, (Awards DGE-1315138 and DGE-1842473).}

\bibliographystyle{abbrv-doi}

\bibliography{main.bib}
\end{document}